\documentstyle[12pt,a4wide]{article}
\input epsf
\newcommand{\be}{\begin{equation}}
\newcommand{\ee}{\end{equation}}

\newcommand{\bpsi}{\mbox{\boldmath $\psi$}}

\newcommand{\news}{\setcounter{equation}{0}}
\def\ben{\begin{equation}}
\def\een{\end{equation}}
\def\bea{\begin{eqnarray}}
\def\eea{\end{eqnarray}}
\begin{document}

\title{\vskip -1cm
\bf \large \bf Spinning Skyrmions and the Skyrme Parameters\\[30pt]
\author{Richard A. Battye$^{1}$, Steffen Krusch$^{2}$
and Paul M. Sutcliffe$^{2}$\\[10pt]
\\{\normalsize $^{1}$
{\sl Jodrell Bank Observatory, Macclesfield, Cheshire SK11 9DL U.K.}}
\\{\normalsize {\sl $\&$  School of Physics and Astronomy,
Schuster Laboratory,}}
\\{\normalsize {\sl University of Manchester, Brunswick St,
 Manchester M13 9PL, U.K.}}
\\{\normalsize {\sl Email : rbattye@jb.man.ac.uk}}\\
\\{\normalsize $^{2}$  {\sl Institute of Mathematics,
University of Kent, Canterbury, CT2 7NF, U.K.}}\\
{\normalsize{\sl Email : S.Krusch@kent.ac.uk}}\\
{\normalsize{\sl Email : P.M.Sutcliffe@kent.ac.uk}}\\[0.5cm]}}
\date{July 2005}
\maketitle

\begin{abstract}
The traditional approach to fixing the parameters of the Skyrme
model requires the energy of a spinning Skyrmion to reproduce
the nucleon and delta masses. The standard Skyrme parameters, which are
used almost exclusively, fix the pion mass to its experimental value
and fit the two remaining Skyrme parameters by approximating the
spinning Skyrmion as a rigid body. In this paper we remove the rigid
body approximation and perform numerical calculations which allow
the spinning Skyrmion to deform and break spherical symmetry.
The results show that if the pion mass is set to its experimental value
then the nucleon and delta masses can not be reproduced for any values
of the Skyrme parameters; the commonly used Skyrme parameters are simply
an artifact of the rigid body approximation. However, if the pion mass
is taken to be substantially larger than its experimental value then
the nucleon and delta masses can be reproduced. This result has a 
significant effect on the structure of multi-Skyrmions.
\end{abstract}

\newpage

\section{Introduction}\news
The Skyrme model \cite{Sk} is a nonlinear theory of pions which is
an approximate, low energy effective theory of quantum chromodynamics,
obtained in the limit of a large number of quark colours \cite{Wi}.
Skyrmions are topological soliton solutions of the model and
are candidates for an effective description of nuclei, with an identification
between soliton and baryon numbers.

The Lagrangian of the Skyrme model contains only three free parameters;
two of these set the energy and length units and the third corresponds
to the (tree-level) pion mass. In Ref.\cite{ANW} the energy and length
units were calculated by fitting to the masses of the nucleon and delta
resonance assuming massless pions, and in Ref.\cite{AN} this calculation
was repeated using the experimental value for the pion mass. This set
of parameters is invariably used in the study of the Skyrme model,
and so we shall refer to them as the standard values.

Recently \cite{BS10} the properties of classical multi-Skyrmions have been
investigated for a range of Skyrme parameters and it has been found that
important qualitative differences arise as the Skyrme parameters are
varied. In particular, setting the pion mass to be substantially
larger than its experimental value yields multi-Skyrmions whose qualitative
features appear to be closer to those of real nuclei. For example, the size of
nuclei scales with mass number $A$ like $A^{1/3}$ rather than like $\sqrt{A}$,
and the stability properties of small nuclei with $A=5$ and $A=8$ 
also seem more realistic. These results provide motivation to re-examine the
original work in Refs.\cite{ANW,AN} where the standard Skyrme parameters were
first calculated.

The approach of Refs.\cite{ANW,AN} involves the zero-mode quantization
of a single Skyrmion as a rigid body.
This is effectively a study of spinning Skyrmions which assumes
that a Skyrmion does not deform as it spins. Early papers
\cite{BR,RSWW} pointed out the limitations of this approximation
and improved upon it by allowing the Skyrmion to deform within
a spherically symmetric hedgehog ansatz. Including only this deformation
already reveals that with the experimental value of the pion mass
there are now no values of the Skyrme parameters that can fit the masses
of both the nucleon and delta. Thus, the standard Skyrme parameters are simply
an artifact of the rigid body approximation.

The need to take into account the deformation of a spinning Skyrmion,
and in particular to allow deformations which break spherical symmetry,
has been noted by several authors \cite{Sc,DHM} using a range of
different physical perspectives. In this paper we address this issue
by performing numerical computations of spinning Skyrmions in the full
field theory assuming only an axial symmetry. We compute the energies
of spinning Skyrmions for a range of Skyrme parameters and pion masses,
and find that the nucleon and delta masses can be fit to the experimental
values only if the pion mass is taken to be larger than twice the
experimental value. A larger value for the tree-level pion mass  
is not necessarily in conflict with the smaller experimental value,
as we discuss in the next section.

\section{Spinning Skyrmions}\news
The field of the Skyrme model \cite{Sk} is an $SU(2)$-valued scalar
$U.$ It is convenient to introduce the $su(2)$-valued current
 $R_\mu=(\partial_\mu U)U^\dagger$ and write the Lagrangian as
\be L=\int \left\{-\frac{F_\pi^2}{16}\mbox{Tr}(R_\mu
R^\mu)+\frac{1}{32e^2}
\mbox{Tr}([R_\mu,R_\nu][R^\mu,R^\nu])+\frac{m_\pi^2F_\pi^2}{8}\mbox{Tr}(U-1)
\right\} \, d^3x. \label{skylag} \ee Here $F_\pi$, $e$ and $m_\pi$
are parameters, whose values are fixed by comparison with
experimental data. 
$F_\pi$ may be interpreted as the pion decay
constant, $e$ is a dimensionless constant and $m_\pi$ 
has the interpretation of the tree-level pion mass 
(we use units in which $\hbar$ is one). The experimental values for the
pion decay constant and pion mass are $F_\pi=186\mbox{ MeV}$
and $m_\pi=138\mbox{ MeV}$ respectively.

The approach of Refs.\cite{ANW,AN} is to fix $F_\pi$ and $e$ by fitting
the energies of a quantized Skyrmion to the masses of the nucleon
and delta resonance, assuming either massless pions \cite{ANW} or the
experimental value for the pion mass \cite{AN}, yielding
the standard values $m_\pi=138\mbox{ MeV},$
$F_\pi=108\mbox{ MeV}$ and $e=4.84.$ 
Note that this value of $F_\pi$ is substantially lower than
the experimental value, but
if $F_\pi$ and $m_\pi$ are both fixed to the experimental values 
 then the only free parameter is
$e,$ and this does not allow a simultaneous fit for both the nucleon
and delta masses. 

The quantization of the Skyrme model is difficult due to the fact that
it is a non-renormalizable field theory. One optimistic approach is
that these effects can be partially modeled by a renormalization of
the Skyrme parameters, in which case $F_\pi$ and $m_\pi$ could be 
interpreted as the renormalized pion decay constant and the 
renormalized pion mass. Therefore values which differ from the
experimental values are not necessarily a contradiction.
Furthermore, we are mainly interested in reproducing the
properties of nuclei for all nucleon numbers and are willing to 
place less emphasis on matching the pion physics if  
necessary.
In this paper we shall treat the three Skyrme parameters 
$F_\pi,$ $e$ and $m_\pi$ all as free parameters, and indeed
we find that neither $F_\pi$ nor $m_\pi$ can be set to their experimental
value if the nucleon and delta masses are to be fit, once the deformation
of a spinning Skyrmion is taken into account.

It is useful to rescale the spacetime coordinates by
$x_\mu\mapsto 2x_\mu /({eF_\pi})$ so that the relative coefficient
between the first two terms in (\ref{skylag}) is independent of the Skyrme
parameters. After applying this rescaling the Lagrangian becomes
\be
L=\frac{F_\pi}{4e}\int \left\{-\frac{1}{2}\mbox{Tr}(R_\mu R^\mu)
+\frac{1}{16}
\mbox{Tr}([R_\mu,R_\nu][R^\mu,R^\nu])+m^2\mbox{Tr}(U-1)
\right\} \, d^3x
\label{skylag2}
\ee
where we have introduced the rescaled pion mass $m=2m_\pi/(F_\pi e).$
From this form of the Lagrangian it is now clear that the parameter
combinations
$F_\pi/(4e)$ and $2/(eF_\pi)$ merely determine the energy and length units
 of classical static Skyrmions. However, the rescaled pion mass $m$
can not be scaled away and so can have an effect on Skyrmion structure.
A recent study \cite{BS10} has demonstrated that
significant qualitative differences arise for multi-Skyrmions
as the parameter $m$ is varied, so the determination of the parameters
of the Skyrme model (and in particular $m$) is important.
Note that, in the quantization of a Skyrmion we must consider its spin, and
then the parameters $F_\pi$ and $e$ can not simply be scaled away.

For a classical static Skyrmion the energy derived from the Lagrangian
(\ref{skylag2}) is
 \be
E_0=\frac{F_\pi}{4e}\int \left\{-{1 \over 2}\mbox{Tr}(R_iR_i)-{1 \over 16}
\mbox{Tr}([R_i,R_j][R_i,R_j])+m^2\mbox{Tr}(1-U)\right\} \, d^3x\,.
\label{skyenergy0}
\ee
The standard approach of Refs.\cite{ANW,AN} is to quantize the rotational
zero modes of the Skyrmion as a rigid body, which yields multiplets
with equal spin and isospin in each multiplet. The details are as follows.

In spherical polar coordinates ($r,\theta,\phi$) the ansatz for a
hedgehog Skyrmion rigidly rotating around the $\phi$-axis is
\be
U=\cos f+i\sin f(\tau_3 \cos\theta +\sin\theta
(\tau_1\cos(\phi+\omega t)+\tau_2\sin(\phi+\omega t)))
\ee
where $\tau_i$ denote the Pauli matrices, $f(r)$ is the
radial profile function (with boundary conditions
$f(0)=\pi$ and $f(\infty)=0$) and $\omega$ is the rotation
frequency in the rescaled coordinates; note that the
rotation frequency in physical units is given by $\Omega=\omega eF_\pi/2.$

Substituting this ansatz into the Lagrangian (\ref{skylag2}) gives
\be
L=\frac{1}{2}\Lambda \Omega^2 - E_0,
\label{lag2}
\ee
where $E_0$ is the static energy
\be
E_0=\frac{\pi F_\pi}{e}\int_0^{\infty}\left\{
r^2f'^2+2\sin^2f \, (1+f'^2)+\frac{\sin^4f}{r^2}
+2m^2r^2(1-\cos f)\right\} \, dr,
\label{hhenergy}
\ee
and $\Lambda$ is the moment of inertia
\be
\Lambda=\frac{16\pi}{3e^3F_\pi}\int_0^\infty\left\{
\sin^2f\, \left(r^2(1+f'^2)+\sin^2f\right)\right\} \, dr\,.
\ee
The equation for the profile function which follows from the Lagrangian
(\ref{lag2}) can equivalently be obtained my minimizing the total energy
\be
E=E_0+\frac{J^2}{2\Lambda},
\label{energy2}
\ee
where $J=\Omega \Lambda$ is the spin, which is a conserved quantity.

The classical Skyrmion is quantized within the Bohr framework by requiring
the spin to be quantized as $J^2=j(j+1),$ where $j$
is the spin quantum number taking values $j=\frac{1}{2}$ for the nucleon
and $j=\frac{3}{2}$ for the delta. Thus, within the classical picture,
the nucleon and delta are simply spinning Skyrmions with a particular
rotation frequency.

In the above discussion we have assumed that the spinning Skyrmion
remains spherically symmetric. In the treatment of
Refs.\cite{ANW,AN} a further approximation is employed, namely that
the Skyrmion does not deform at all when it spins. In this approach
the profile function is not determined by minimization of the total
energy (\ref{energy2}), but rather by minimization of only the static
energy $E_0.$ The additional contribution to the energy due to the
spin is then calculated given the static energy minimizing profile
function. A crucial point is that this rigid body approximation
allows any value of the spin to be obtained for a spinning Skyrmion,
since one simply
 sets the rotation frequency to be $\Omega=J/\Lambda,$ where $\Lambda$
is the moment of inertia of the static Skyrmion.

\begin{figure}[ht]
\begin{center}
\leavevmode \vskip -3.5cm \epsfxsize=15cm\epsffile{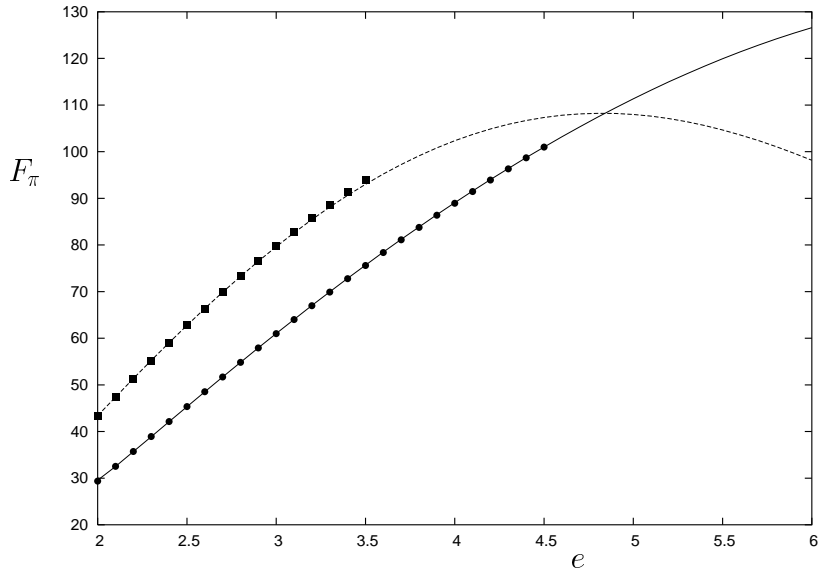} \vskip
-10cm \caption{For $m_\pi=138\mbox{ MeV}$, the graph displays, as a
function of the Skyrme parameter $e$, the value of $F_\pi$ required
to fit the nucleon mass (solid curve for rigid body approximation
and circles for axial deformation) and delta mass (dashed curve for
rigid body approximation and squares for axial deformation).
Note that the data for the axial deformation of the nucleon and 
delta do not cross.}
\label{fig-one} \vskip 0cm
\end{center}
\end{figure}

Setting the pion mass to its experimental value ($m_\pi=138\mbox{
MeV}$) leaves the remaining parameters $F_\pi$ and $e$ to be
determined by fixing the energy of the $j=\frac{1}{2}$ and
$j=\frac{3}{2}$ spinning Skyrmions to the masses of the nucleon
($M_N=939\mbox{ MeV}$) and delta ($M_\Delta=1232\mbox{ MeV}$). In
Fig.~\ref{fig-one} we plot, for a range of $e,$ the value of $F_\pi$
required to fit the nucleon mass (solid curve) and the value of
$F_\pi$ required to fit the delta mass (dashed curve) using the
rigid body approximation. These two curves cross at the value
$e=4.84$ with $F_\pi= 108 \mbox{ MeV},$ which are the standard
parameter values obtained in Ref.\cite{AN} and produce a rescaled
pion mass $m=0.526.$

The limitations of the rigid body approximation were highlighted in
the papers \cite{BR,RSWW} and a first improvement was made by allowing
the Skyrmion to deform within a spherically symmetric hedgehog
ansatz. In this approach the profile function $f(r)$ is determined
by minimization of the total energy (\ref{energy2}). An analysis of
the ordinary differential equation for the profile function reveals
that a localized finite energy solution exists only if the rotation
frequency satisfies the constraint $\Omega^2\le \frac{3}{2}m_\pi^2.$
Furthermore, at the maximal rotation frequency
$\Omega=\sqrt{\frac{3}{2}}m_\pi$ the moment of inertia $\Lambda$ is
finite so there is an upper bound on the spin $J.$ As the Skyrme
parameter $e$ is increased there is a point at which the required
spin for the delta (or nucleon) exceeds the upper bound, so there is
no spinning Skyrmion that describes the delta (or nucleon). In
Refs.~\cite{BR,RSWW} it was found that with the experimental value of
the pion mass the standard values for $e$ and $F_\pi$ are in the
region for which there is no spinning Skyrmion solution for the
delta, so the standard Skyrme parameters are simply an artifact of
the rigid body approximation and there are no Skyrme parameters that
can be chosen to fit the masses of both the nucleon and delta.

Physically one expects the constraint on the spin to be $\Omega^2\le
m_\pi^2,$ since this expresses the fact that the Skyrmion can only
spin at a frequency upto the pion mass before it begins to radiate
pions. The extra factor of $\sqrt{\frac{3}{2}}$ in the constraint
for the spinning hedgehog field reflects the fact that the radiation
is assumed to be a spherical wave in the hedgehog approximation. A
consistent ansatz for a spinning Skyrmion can have at most an axial
symmetry and will radiate differently in the directions orthogonal
and parallel to the symmetry axis. Thus, although the deformed
hedgehog approximation captures the qualitative feature that there
is a maximal spin, the quantitative details require the computation
of spinning Skyrmions with axial symmetry.

A limited numerical study (using small grids) of axially
symmetric spinning Skyrmions has been performed \cite{WWS}, but for only
one set of Skyrme parameters, so the dependence on the Skyrme
parameters that we wish to address in this study has not been investigated.

Introducing cylindrical polar coordinates $\rho,\chi,z$ the ansatz
for an axially symmetric spinning Skyrmion is a simple generalization
of the one introduced in Ref.\cite{KS} and is given by
\be
U=\psi_3
+i\psi_2\tau_3+i\psi_1(\tau_1\cos(\chi+\omega
t)+\tau_2\sin(\chi+\omega t))
\label{axial}
\ee
where $\bpsi(\rho,z)=(\psi_1,\psi_2,\psi_3)$ is a three-component unit
vector that is independent of $\chi.$
The boundary conditions on $\bpsi$ are that
$\bpsi\rightarrow (0,0,1)$ as $\rho^2+z^2\rightarrow \infty,$
and on the symmetry axis $\rho=0$ we require $\psi_1=0$ and
$\partial_\rho\psi_2=\partial_\rho\psi_3=0.$

With the axial ansatz (\ref{axial}) the expressions for the static
energy and moment of inertia become 
\be E_0=\frac{\pi F_\pi}{2e}\int
\{ (\partial_\rho\bpsi\cdot\partial_\rho \bpsi +
\partial_z\bpsi\cdot\partial_z \bpsi) (1+\frac{\psi_1^2}{\rho^2})
+|\partial_\rho\bpsi \times \partial_z\bpsi|^2
+\frac{\psi_1^2}{\rho^2} +2m^2(1-\psi_3) \}\ \rho\,d\rho\,dz
\label{axialenergy} 
\ee 
\be \Lambda=\frac{4\pi}{e^3F_\pi} \int
\{\psi_1^2 (\partial_\rho\bpsi\cdot\partial_\rho \bpsi +
\partial_z\bpsi\cdot\partial_z \bpsi+1) \}\ \rho\,d\rho\,dz\,.
\label{axialinertia} \ee 
For a given spin $J=\sqrt{j(j+1)}$ the
configuration which minimizes the total energy (\ref{energy2}) needs
to be computed, with $E_0$ and $\Lambda$ given by
(\ref{axialenergy}) and (\ref{axialinertia}). This minimization is
performed numerically using a simulated annealing algorithm on a
grid in the $(\rho,z)$-plane containing $250\times 500$ grid points
and a lattice spacing of 0.06.

With the pion mass set to its experimental value
of $m_\pi=138\mbox{ MeV},$ we calculate, for a range of the Skyrme
parameter $e,$ the value of $F_\pi$ required to match the energy
of a $j=\frac{1}{2}$ spinning Skyrmion to the nucleon mass and
also the value of $F_\pi$ required to match the $j=\frac{3}{2}$ spinning
Skyrmion energy to the delta mass.

The results of the simulated annealing minimization are displayed in
Fig.~\ref{fig-one}, where circles denote the values required for the
nucleon fit and squares denote the delta fit. Note that the results
of the axial calculation (circles and squares) are extremely close
to those of the rigid body approximation (solid and dashed curves),
indicating that for this value of the pion mass there is little
deformation of the spinning Skyrmion. However, the crucial point is
that the spinning Skyrmion exists only if the parameter $e$ is less
than a critical value, which is why the axial data in
Fig.~\ref{fig-one} terminates. At the termination point, the rotation
frequency $\Omega,$ which we calculate as $J/\Lambda,$ is equal to
the pion mass $m_\pi$ and therefore the Skyrmion has attained its
maximal value of the spin. If $e$ is increased beyond this value the
Skyrmion can not be spun fast enough to reach the required spin $J$
with the new parameters, and so the required spinning Skyrmion
solution does not exist. Clearly, since the spin of the delta is
greater than that of the nucleon the termination point for the delta
is at a lower value of $e$ than for the nucleon.

\begin{figure}[ht]
\begin{center}
\leavevmode \vskip -3.5cm \epsfxsize=15cm\epsffile{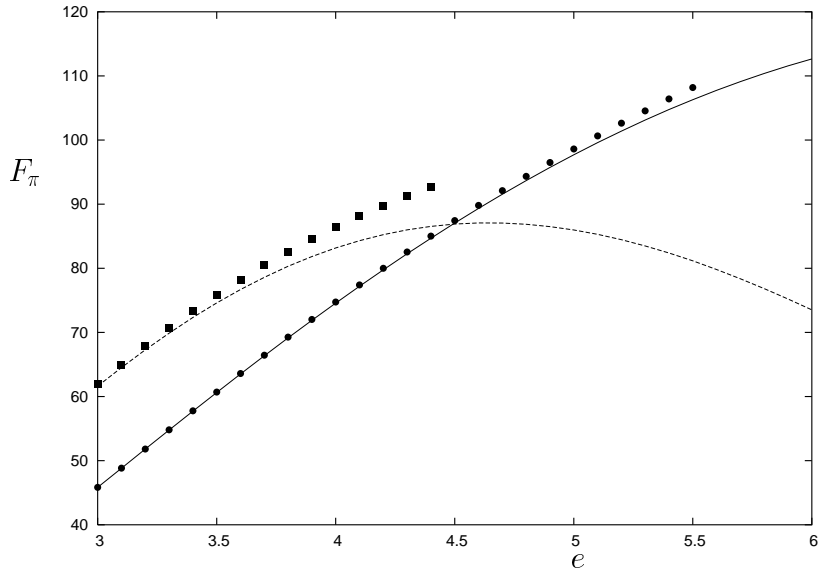} \vskip
-10cm \caption{As Fig.~\ref{fig-one} but with $m_\pi=276\mbox{
MeV}$. Note that the data for the axial deformation of the nucleon and 
delta still do not cross.} \label{fig-two} \vskip 0cm
\end{center}
\end{figure}

\begin{figure}[ht]
\begin{center}
\leavevmode \vskip -3.5cm \epsfxsize=15cm\epsffile{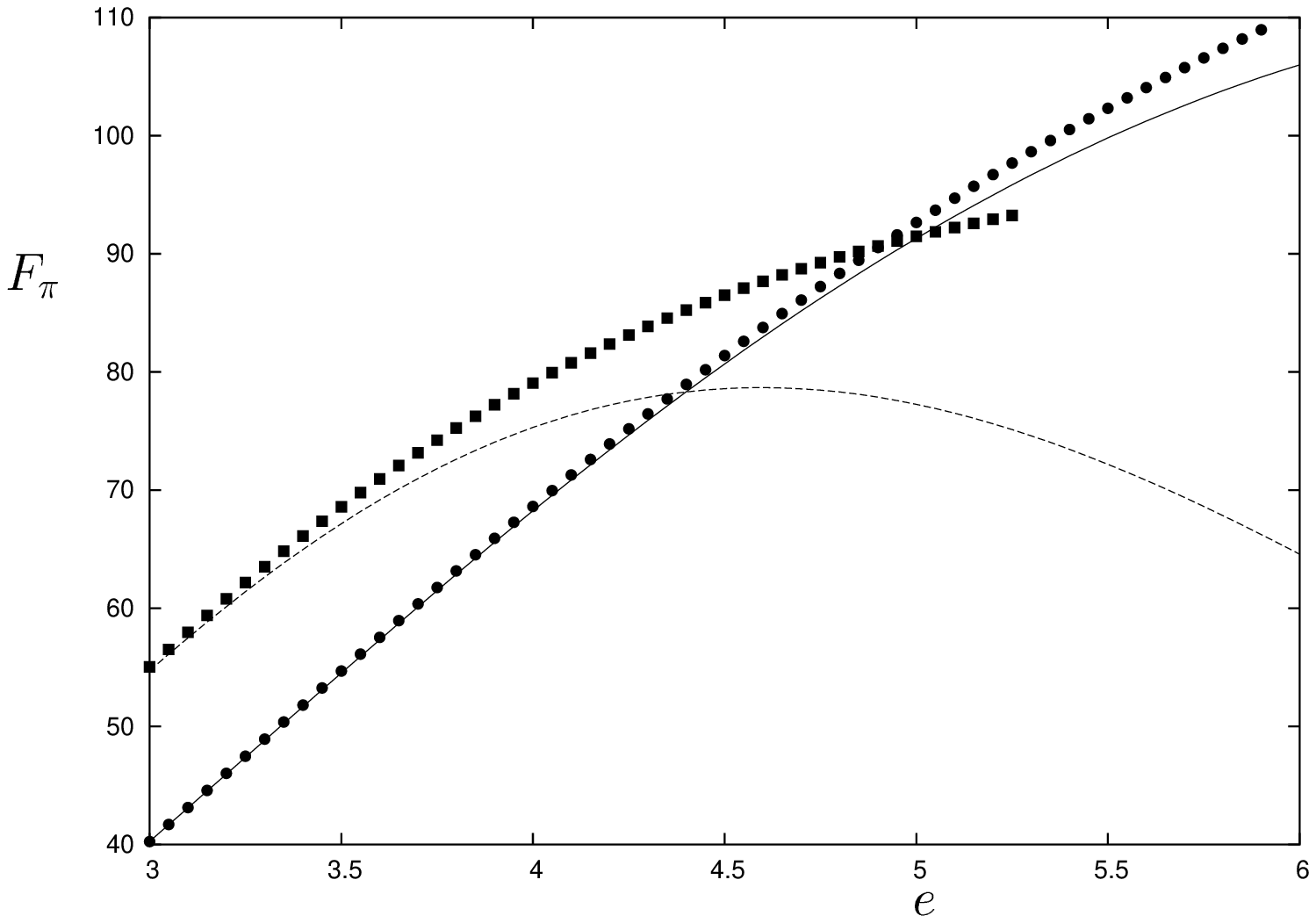}
\vskip -10cm \caption{As Fig.~\ref{fig-one} but with
$m_\pi=345\mbox{ MeV}$.
Note that the data for the axial deformation of the nucleon and 
delta now cross.} \label{fig-three} \vskip 0cm
\end{center}
\end{figure}

A linearization of the $\psi_1$ field equation yields
\be
\partial_i\partial_i\psi_1-(m^2-\omega^2)\psi_1=0
\ee so a finite energy solution exists only if the effective mass is
non-negative ie. $\omega^2\le m^2,$ or in physical units
$\Omega^2\le m_\pi^2,$ which is the criterion we have used to
identify the termination point of the spinning Skyrmion. On a finite
numerical grid all configurations have finite energy so care has to
be taken in determining the termination point. It would be easy to
mistakenly compute a configuration with a rotation frequency which
is apparently greater than $m_\pi,$ but one can check that for such
a configuration the computed energy is not independent of the size
of the numerical grid, which signals that this is not a finite energy
solution.

The crucial point about the axial data in Fig.~\ref{fig-one} is that
there is no crossing point of the nucleon and delta data, so there
are no values of $e$ and $F_\pi$ that fit both the nucleon and delta
masses. The crossing point within the rigid body approximation is
well beyond the termination points of the axial data, so the
standard values are an artifact of the approximation and do not
correspond to any spinning Skyrmion solutions.

If the nucleon and delta masses are to be matched to the energies of
spinning Skyrmions then the only possibility is to increase the pion
mass beyond its experimental value. Increasing $m_\pi$ allows the
Skyrmion to spin at a greater frequency, so larger values of $e$ can
be attained, and it may be possible that both masses can be fit
simultaneously.

It is of interest to note that several other studies, from different
perspectives, have indicated that a pion mass larger than the
experimental value yields Skyrmion results that agree better with
experimental data. For example, studies of the Roper resonance
\cite{BrN} and comparisons between vibrational frequencies of 
multi-Skyrmions and nuclear gamma ray spectra \cite{BBT}, 
produce improved results with a larger pion mass.

In Fig.~\ref{fig-two} we display (using the same notation as in
Fig.~\ref{fig-one}) the results of both the rigid body and axial
calculations with the pion mass set at twice the experimental value
$m_\pi=276\mbox{ MeV}.$ Note that now the axial computations show 
some differences from the rigid body approximation when the delta
spins close to its allowed limit, indicating an increased
deformation from the static Skyrmion. Although the nucleon and delta
data are now closer to an intersection, they still do not cross, so
again there are no values of the Skyrme parameters to fit both
masses.

 An examination of the spinning Skyrmion configuration
reveals that most of the deformation can be captured within a
hedgehog ansatz, so the leading correction is associated with a
deformation of the size of the Skyrmion. However, if the deformed
hedgehog approximation of Refs.\cite{Sc,DHM} is implemented the
nucleon and delta curves actually cross for this value of the pion
mass, as the termination points are not correctly predicted.

Increasing the pion mass a little more, to two and a half times the
experimental value $m_\pi=345\mbox{ MeV}$, produces the data shown
in Fig.~\ref{fig-three}. Now the axial delta data is substantially
different from the rigid body approximation, and the nucleon and
delta curves intersect. With this value of the pion mass the Skyrme
parameters can be determined by fitting to the nucleon and delta
masses and the parameters are found to be $e=4.90$ and
$F_\pi=90.5.$ This corresponds to a rescaled pion mass of $m=1.56,$
which is remarkably close to the value required in the study of the
Roper resonance \cite{BrN}.

It is interesting that the above values of $e$ and $F_\pi$ are
similar to those of the standard parameters, but now require a pion
mass which is more than twice the experimental value. There is
clearly a critical pion mass in the range $276\mbox{ MeV }\le m_\pi
\le 345\mbox{ MeV}$ above which a fit to the nucleon and delta
masses is possible and below which it is not, but we have not
computed this since 
it requires a substantial number of simulations
and is probably not the best way to fit the Skyrme parameters. The values of
$e$ and $F_\pi$ at the crossing point (when it exists) appear to be
fairly stable to changes in the pion mass, so the above set of
values seems a suitable set for further investigations.

In comparing our results with previous studies, we note that
Skyrmions spin rather slowly in the sense that the deformation from
the static Skyrmion is quite small even when the rotational
frequency takes its maximal value. Only for a large pion mass (which
sets the maximal rotation frequency) and then only for the delta, is
a significant deformation found. This agrees with the findings of
Ref.\cite{WWS} in the sense that there it was also found that substantial
axial deformations occur only for large values of the spin.
However, we find that such solutions can not exist as finite
energy spinning Skyrmions unless the Skyrme parameters are
substantially different from the standard values. 

Our results should be contrasted with similar studies on planar baby
Skyrmions \cite{PSZ2}, where the important difference is that the
moment of inertia of a baby Skyrmion is infinite as the maximal
rotation frequency is reached, thus allowing any spin to be
attained. In this case the deformation close to the allowed maximal
spin is substantial, as one would expect from a diverging moment of
inertia.

\section{Conclusion}\news
In this paper we have used an axially symmetric numerical code to
compute the energies of spinning Skyrmions and have determined the
parameters of the Skyrme model to fit these energies to the masses
of the nucleon and delta. We have described how the standard Skyrme
parameters are obtained in a region which is not physical and are an
artifact of the rigid body approximation. Moreover, we have shown
that the nucleon and delta masses can be matched only if the pion
mass is set at more than twice its experimental value. This result
supports recent work \cite{BS10} on multi-Skyrmions, which finds
that setting the pion mass to be substantially larger than its
experimental value yields multi-Skyrmions whose qualitative features
are closer to those of real nuclei.

A valid criticism of fitting the Skyrme parameters to the nucleon
and delta masses is that the delta is an unstable resonance, which
nevertheless must be modeled as a stable spinning Skyrmion. 
Note that using the rigid body approximation,
or the axial ansatz with a rotation frequency above the pion mass,
does not model the delta as an unstable solution since there is
simply no solution, rather than an unstable one. It would therefore
seem that a better approach to fixing the Skyrme parameters is to
use properties of multi-Skyrmions, and these appear to favour a
larger pion mass \cite{BS10}. 
The main result of the present paper
is to conclude that a large pion mass is not incompatible with studies of the
single Skyrmion, and indeed the nucleon and delta calculations
support this view.

Finally, the parameter $m_\pi$ is interpreted as the pion mass but
there are all kinds of quantization issues beyond the zero-mode
quantization, such as the calculation of the Casimir energy \cite{MeWa},
which are difficult to resolve due to the fact that the Skyrme model
is a non-renormalizable field theory. 
As we have mentioned earlier, one optimistic approach is
that these effects can be partially modeled by a renormalization of
the Skyrme parameters, in which case $m_\pi$ could be interpreted as
a renormalized pion mass, and therefore a larger value is not
necessarily in conflict with the smaller experimental value.

\section*{Acknowledgements}
Many thanks to Juan Ponciano, Bernd Schroers and Nick Manton for useful
discussions. This work was supported by the PPARC special purpose
grant ``Classical Lattice Field Theory''. SK acknowledges the EPSRC
for a postdoctoral fellowship GR/S29478/01.

\end{document}